\documentclass[a4paper, 12pt]{article}
\usepackage{tikz} 
\title{The Strength of Varying Tie Strength}
\author{}
\author{Jeroen Bruggeman\thanks{Department of Sociology and Anthropology, 
University of Amsterdam, Oudezijds Achterburgwal 185, 1012 DK Amsterdam, the Netherlands. Email: j.p.bruggeman@uva.nl. Thank you to Sinan Aral,  Douglas R. White, Nina O'Brien, Joe Labianca, Stephanie Steinmetz,  Emily Miltenburg, Herman van de Werfhorst, an anonymous reviewer, and participants at the Sunbelt social network conference in Hamburg.}}
\date{\today}

\begin{document}
\maketitle

\begin{abstract} ``The Strength of Weak Ties" argument (Granovetter 1973) says that the most valuable information is best collected through bridging ties with other social circles than one's own, and that those ties tend to be weak. Aral and Van Alstyne (2011) added that to access complex information, actors need strong ties (``high bandwidth") instead. These insights I integrate and generalize by pointing at actors' benefits and costs. Weak ties are well-suited for relatively simple information at low costs, whereas for complex information, the best outcomes are expected for those actors who vary their bandwidths along with the value of information accessed. To support my claim I use all patents in the USA (two million) over the period 1975---1999. \end{abstract}

\section*{Introduction}
How do people get valuable information\footnote{Depending on the context, the value of information can be attributed to its relevance, accuracy, reliability, novelty, scope, timing, rareness, legitimacy, or a combination of (some of) these factors, in order to achieve a (more) beneficial outcome.} to solve their problems and to satisfy their needs? Sometimes they can get it from their own experience or intelligence. Most of the times, however, they strongly rely on their network as a radar and filter, compare what they hear and see (and read, in modern societies), and attempt to combine useful information into solutions \cite{burt92}. To make beneficial comparisons and combinations, it turns out that, in general, \emph{diverse} information is key \cite{page07}. 

When studying information search from a network perspective, detailed knowledge about the content of ties is usually lacking; it is a challenge to model information diversity in a general way, and predict beneficial use of it in a broad range of fields. Would network diversity, i.e.~a focal node's connections to mutually disconnected nodes, be sufficient, or would additional indicators be necessary? A large portion of the literature confirms that for ego (a focal actor) to access diverse information, (s)he should broker across ``structural holes" between alters, or information sources in general, in different social groups \cite{burt04,stovel12}. Within groups, people are highly clustered; this clustering fosters mutual comprehension and coordination. But continual interactions render shared information relatively homogeneous. Rather, as the most valuable information is heterogeneous, clustering constrains access to it \cite{granovetter73}. Moreover, within-group ties are relatively strong with regards to engagement and time spent, whereas between-group ties tend to be weak \cite{onnela07}. Consequently, ``average" brokers mostly use weak ties to access diverse information across different groups. 

In a recent paper, Aral and Van Alstyne \cite{aralvan11} proposed, by contrast, the notion that for ego to obtain diverse but complex or rapidly changing information, (s)he must use strong ties, called high \emph{bandwidth} in those contexts, even if this implies a loss of some structural holes being brokered. Although strong ties' benefits have been noticed in previous studies \cite{uzzi97,hansen99}, Aral and Van Alstyne support the trade-off they discovered by an impressive data set collected from an executive recruiting firm, as it exceptionally contained not only network ties but also the content of these ties. Among their findings, the authors showed that information diversity is indeed enhanced by network diversity. This finding supports earlier studies in which tie content was largely unknown. In their case study, by helping to shorten project duration, efficiency was the benefit of diverse information. 

In most cases, however, it is unlikely that the information value is the same across sources. One wonders, then, if an optimal average bandwidth could predict the highest benefits. By taking both benefits and costs into account, I attempt to integrate and generalize the weak tie and high bandwidth theses. Relatively simple information can be accessed through weak ties at low transmission, processing, and tie maintenance costs, whereas complex or rapidly changing information requires high bandwidth and effort, all else being equal. However, to access complex information, actors should avoid spending large resources on low value information, and will benefit more if they focus on the best sources they can get access to, given cognitive and institutional limitations. For weak ties to low value sources, in contrast, costs are low and higher exploration risks can be taken. In sum, if bounded rational actors search for simple information, network diversity and weak ties are their best options, but once they have to trade-off network diversity for bandwidth, when information is complex or rapidly changing, the best outcomes are expected for those who vary their bandwidths along with the quality of their sources. This effect of bandwidth variation, then, is the hypothesis to be tested in this paper.

In fields, such as science, technology and in knowledge-intensive industries, where valuable information is complex, actors have to invest time and effort in certain sources, oral or written, and achieve a skillful command of the knowledge they will have acquired through them. To successfully broker and cross-fertilize complex information, accessing those sources must therefore be accompanied by specialization in those sources, which imposes substantial costs. For these actors, specialization is not only a process of accumulating knowledge in a given domain, but also of interrelating their knowledge more densely such that they may discover shortcuts and workarounds. In a service industry, for example, this means that employees have to acquaint themselves with their colleague's skills, knowledge and personalities, their clients' wishes and idiosyncrasies, and learn effective solutions to the problems at hand. Clearly, for obvious cognitive limitations, nobody can specialize in a great many alters or sources simultaneously; this implies a diversity-bandwidth trade-off. This conclusion also holds true for collectives, such as organizations, even though they are able to process more information than individuals.

However, we know from science studies that ``standing on the shoulders of giants," i.e.~using the best sources available, has strong positive effects, both on individuals' careers and on the accumulation of public knowledge \cite{bornmann10}. Furthermore, successful PhD students often have good supervisors \cite{malmgren10}; famous philosophers owe part of their success to being connected to other famous philosophers \cite{collins98}; and, successful innovations build on prior successful ideas \cite{carnabuci10}. In all these studies, information value varies across sources, and obtainable benefits vary with them, which corroborates my bandwidth variation thesis. Because none of these studies is based on a research design specifically targeted toward testing it, I will test it here using a longitudinal data set of patent citations. In the pertaining field of technology, the information in patents is complex (although not rapidly changing), and people have to invest considerable time and effort to master the sources they tap. 

In the next section, I describe the data and the network measures. Subsequently, I present the results, and discuss them in the final section.

\section*{Data and measures}
For the empirical test I use publicly available data on a network in which the nodes are ``invisible colleges" of inventors ($n = 417$), at the aggregate level, and which incorporate all patents in the USA (two million) over the period 1975---1999 (USPTO).\footnote{The data were harvested in 2001 from \texttt{http://www.uspto.gov/}. I exclude from the initial 418 domains one inactive one. Variable values of temporarily inactive domains were coded ``NA" (non available) to prevent taking the logarithm of zero later on.} The administrative units corresponding to these colleges of inventors are technology domains wherein patents are categorized. These units are both stable and non-overlapping over the period of observation, while other units are non-stable, overlapping, or both \cite{henderson05}. Just like scientific citations, patent citations (as directed ties in the network) represent knowledge flows, and explicate which ideas have been (re)combined into a new idea.  The period of observation is partitioned into non-overlapping sub-periods, lasting five years each, consistent with other studies using patent data \cite{podolny96}. Over the 25 years of observation, the density of the network ($m/n^2$, for a network with loops) gradually increased from 0.217 in the first period to 0.395 in the last,  while the average path length (concatenation of ties from a node  to node another) shrank from 1.83 to 1.62.

Actors can self-specialize by re-using knowledge produced earlier. At
the level of a domain, self-specialization means that a myriad of individuals within the domain cite each other and sometimes themselves. Self-specializing individuals build upon earlier experiences, which they integrate
with their current experiences or with others' information. In a network,
self-specialization is indicated by a tie from a node to itself (reflexive arc, or loop).
For technology domains, self-specialization ties are the strongest on average,
and increased from 1508 to 7287 citations over the period of observation, compared
to 1086 to 6722 for dyadic ties, respectively, in the range from 0 to 79337. Further details of these data have been presented and discussed elsewhere \cite{carnabuci09}. 

To measure nodes' \emph{brokerage} (network diversity), I use betweenness centrality \cite{brandes08,freeman77}; see Appendix. It is independent from tie strength variation that should be measured separately. Whereas betweenness normally takes into account all shortest paths from here to there through a focal node, recent studies have shown that information further away than ego's direct connections does not contribute to ego's brokerage opportunities \cite{burt07,aral07}. In line with these findings I truncate shortest paths longer than two ties, and only count structural holes between pairs of unconnected nodes in direct contact with ego. We may call this measure between-two-ness, 
to contrast it with the earlier notion that boiled down to between-everybody-ness. Obviously, a tie between two alters closes the structural hole for ego, renders alters' information more homogeneous, and does not contribute to ego's brokerage. Yet heterogeneity is also reduced by multiple indirect contacts between alters \cite{moodywhite}, which at the same time diminish exclusive access to the structural hole. Accordingly, the value of a structural hole is reduced by the number of nodes that have access to it (i.e., are structurally equivalent with respect to the structural hole). Finally, in citation networks where actors create valuable ideas by (re)combining information they take \emph{from} sources, rather than just being middle (wo)men, I restrict between-two-ness to structural holes among actors' outgoing (citation) ties.  

\emph{Bandwidth} (average tie strength) can be calculated by weighted outdegree (nr.~of citations) divided by outdegree (nr.~of cited domains). For its \emph{variation}, I use an entropy measure \cite{eagle10}, to facilitate comparison with other studies and data.  The following example is meant to foster an intuitive understanding. If you want to know where Bianca bought the book you notice on her table, when you only know that she buys from three different bookshops equally often (her three ties have the same strength), you need some additional information to pin down the particular shop. This information corresponds to an amount of entropy. If, in contrast, your want to know where Abi bought her latest book, knowing that she has one favorite among her three bookstores (one tie is much stronger than the other two), you have a reasonable chance that her book came from there. Your a-priori uncertainty about Abi's bookstore is lower than about Bianca's, and an average search for it requires less additional information, expressed by lower entropy. In the general measure, entropy is highest if all ties are equally strong, and low if the focal node has one or few strong ties, and weak ties with its remainder contacts. The measure is based on proportional tie strengths, $p_{ij}$, to be reached by row-normalizing the adjacency matrix, after ensuring that the arcs point to the direction of citations (or of information asked, in studies of organizations). It is normalized for the number of ties, which have already been taken into account in between-two-ness. For node $i$ with (out)degree $k_i$, normalized tie strength entropy is calculated by

\begin{equation}
     S_{N}(i) = \frac{-\sum_{j=1}^k p_{ij} log(p_{ij})}{log(k_i)}
    \label{entropy}
    \end{equation}

\emph{Information value}, or the quality of its source, is arguably difficult to measure in general, but in the technology and science fields one can measure citation impact \cite{griliches90}. This indicates, crudely, how valuable or relevant others find certain information to be to their own inventions or research, respectively. Although this measure is incomplete and possibly biased as with regards to individual patents, noisy data can be informative at the aggregate level of technology domains \cite{jaffe02}. I include self-citations, because they indicate part of the economic viability of a domain \cite{hall12}. In short, I sum over the columns of the adjacency matrix, $\mathbf{A}$, to create a (column) vector of citation impact, $\mathbf{y}$, for each of the five sub-periods. 

Because it takes time to patent, and to separate cause and effect, the predictors are lagged over one five-year period, $L = t - 1$, where $t$ is an index of time periods and $L$ is the lag. For the response variable, \emph{benefits} obtained by using certain information, I use the focal nodes' current citation impact, because it indicates the economic value for the patent holders \cite{jaffe02}. 

To assess the effect of a node's variation of tie strengths along the value of its sources, I construct a measure of network autocorrelation \cite{leenders02}, in which each weighted outgoing tie is multiplied by the value (citation impact) of the node (domain) being cited.
I row-normalize the adjacency matrix \cite{leenders02}, resulting in a matrix $\mathbf{W}$. My hypothesis can then be formalized as $ \mathbf{y} \propto \mathbf{W}_{L} \mathbf{y}_{L}$. 
From entropy and bandwidth, self-citations are excluded, and the parametrized model is: 

\begin{equation}
\mathbf{y} = \rho \mathbf{W}_{L}\mathbf{y}_{L} + \mathbf{X}_{L}\mathbf{\beta} + \mathbf{\epsilon}. 
\label{statmodel2}
\end{equation}
Because of unobserved heterogeneity, e.g.~variation in R\&D spending across domains and time, Eq.~2 must be expanded to a fixed effect panel model with time dummies. Furthermore, some of the covariates of nodes in ego's network neighborhood might be autocorrelated with $\mathbf{Wy}$, and consequentially, estimates in Eq.~2 could be biased \cite{doweff09}. Therefore I first estimate 
\begin{equation}
\mathbf{Wy} =  \mathbf{W}^{\diamond}_{L}\mathbf{X}_{L}\mathbf{\alpha} + \mathbf{\epsilon}_\diamond, 
\label{statmodel1}
\end{equation} where $\mathbf{W}^{\diamond}$ is obtained from $\mathbf{W}$ by setting the diagonal to zero, and by row-normalizing the remainder ties. It turns out that in Eq.~3, autocorrelated entropy is significant, hence I add it as a control to Eq.~2.
 
For all variables except entropy, the logarithm is used to make their
distributions symmetrical, to straighten their curvilinear relations before computing the correlations (Table 1), and to use them appropriately in the models. 
For the same reasons, entropy is raised to the power of 1.5.

\begin{table}[ht]
\begin{center}
\begin{tabular}{rrrrrr}
  \hline
 & cited & varval & between & bandw & entropy \\ 
  \hline
cited &  & 0.667 & 0.830 & 0.859 & -0.484 \\ 
  varying-with-value &  &  & 0.452 & 0.727 & -0.450 \\ 
  between-two-ness &  &  &  & 0.635 & -0.250 \\ 
  bandwidth &  &  &  &  & -0.666 \\ 
   \hline
mean    & 6699  & 10375 & 127.50  & 17.67 & 0.749  \\
SD      & 10651  & 9819  & 199.74 & 20.01 &  0.082  \\ \hline

\end{tabular}
\caption{\small Correlation matrix; the predictors have their one-period lag. The logarithm is used for all
variables, except entropy that is raised to the power of 1.5. For the means and standard deviations, the raw data are used.}
\end{center}
\end{table}

\section*{Results}
The results are presented in Table~2, with the empty model for the time periods in the first column.   For the effects and models, $P < 0.001$, except for some insignificant effects indicated by a dot.
 
Model Varval shows the effect of varying tie strength with information value, and supports my hypothesis.\footnote{When leaving out self-citations, and substituting $\mathbf{W}^{\diamond}$ for $\mathbf{W}$, the coefficient becomes 0.846 (0.086). When modeling without tie strength altogether, thus ignoring its variation and substituting a binarized (and then row-normalized) matrix for $\mathbf{W}$, the coefficient becomes 0.743 (0.165). In both variations of the Varval model, the overall R-squared is lower, namely 0.705 and 0.687, respectively.} Yet I establish two additional tests:  I examine the effects of average bandwidth and its variation separately, and then put all variables together. Having high bandwidth on average is beneficial (Band), a result that provides strong support for Aral and Van Alstyne's thesis because its confirmation comes from an entirely different field. Low variation of tie strength (high entropy) relates negatively to citation impact (Entro), which shows that also varying bandwidth is beneficial, not only having high bandwidth on average. 

When putting all variables into one model (All), it is clear that brokerage, measured as between-two-ness on outgoing ties,\footnote{When substituting undirected between-two-ness for citation between-two-ness, the significance levels of all effects stay the same, whereas the pertaining coefficient becomes 0.233 (0.019). For the resulting model, $R^2$ = 0.812.} contributes to success, which we could glean from many earlier studies. The remainder variables keep their significance except entropy,  possibly because of multicollinearity (Table~1). When taking these findings together, they do strongly suggest that for the field of technology, my hypothesis is correct.

\begin{table}[ht]
\begin{center}
\begin{tabular}{rrrrrrrr}  \hline
             & Time & Varval  & Band    & Entro       &   All       \\ \hline 
variation with value  
                 &  &  0.888  &         &             &    0.289    \\  
                 &  & (0.059) &         &             &   (0.064)   \\ 
bandwidth        &  &         & 0.839   &             &    0.543    \\ 
                 &  &         & (0.038) &             &   (0.048)   \\ 
entropy          &  &         &         & -2.286      &  $-0.287\bullet$  \\ 
                 &  &         &         & (0.212)     &   (0.222)   \\ 
between-two-ness &  &         &         &             &    0.157    \\
                 &  &         &         &             & (0.016)     \\ 
autocorrelated entropy
                 &  & -0.360  & $0.003\bullet$ & $0.001\bullet$ & $-0.082\bullet$ \\
                 &  &  0.068  & 0.056   &  0.063      &   0.059     \\ \hline
  $R^2$       & 0.674 & 0.731 &  0.773  &  0.710      &   0.814     \\ 
  $F$         & 863   & 664   &  833    &    595      &    657      \\ 
  $n$         & 2085  & 1641  &  1641   & 1638        &   1620      \\ \hline
\end{tabular}
\caption{\small Fixed effect models over four 5-year periods with a one-period lag. For the models and effects in them, $P<0.001$, unless indicated by a dot when insignificant.}
\end{center}
\end{table}

\section*{Discussion and conclusion}
To get valuable information, actors benefit from brokering across diverse, i.e.~disconnected, sources. Aral and Van Alstyne (2011) added that when information is complex or rapidly changing, actors have to trade-off network diversity for bandwidth, while most earlier studies emphasized the strength of weak ties. I integrated and generalized these different findings by looking at costs and benefits: for relatively simple information, network diversity can be accessed through weak ties at low transmission, processing, and tie maintenance costs, whereas for complex or rapidly changing information, when actors have to trade-off network diversity for costly bandwidth, those who vary their bandwidths along with the quality of their sources will have the best outcomes.

In this study, bandwidth variation was assessed at the aggregate level of technology domains, but it is very likely to hold true also for organizations and individuals. Put differently, it seems very unlikely that individuals having their strongest ties to sources that provide irrelevant or false information would outperform those using their strongest ties to acquire the most valuable information instead. 

Bandwidth variation along the value of information was measured by network autocorrelation. Computer simulations \cite{mizruchi08} have demonstrated that in dense networks, such as technology domains, autocorrelation is systematically under-estimated. Therefore, the effect of varying tie strength is even stronger in actuality than estimated here. 

Low variation of bandwidth was shown to be advantageous in a study on British telephone data, which were stripped of content for the sake of privacy \cite{eagle10}. The nodes in that network were postal code-delimited communities, as sources and recipients of information. Tie strength was measured by time spent (volume of calls) by one node calling another node over a period of one month. There, more equally divided attention across sources, indicated by high entropy (Eq.~1), correlated positively with a community-aggregated index of economic welfare ($r = 0.73$). In that study, it is clear that in the large numbers of calls, exchanges of sophisticated information were by far outnumbered by more mundane exchanges. For all those simple subject matters, even valuable ones, strong ties imply redundancy rather than progressive knowledge refinement. Dedicating a great deal of attention to relatively few sources has therefore no advantages, or only briefly, while it precludes people and their communities from getting more valuable information elsewhere. Low variation of bandwidth is therefore advantageous in fields with generically simple information. 

Part of potential information benefits depends on how well actors can integrate new information with their prior knowledge. Actors' absorptive capacity for new information on a given topic is enhanced by specialization in that topic and its sources \cite{levinthal90}. These actors are then better able to notice valuable information amidst noise, including brokerage opportunities that laymen overlook. ``Chance favors the prepared mind," as Louis Pasteur said. Specialization has long term network efficiency as well: once it has reached an adequate level, weaker ties with the pertaining sources will suffice to stay up to date. 

Over a longer period of time, actors may exploit their sources to a point where their potential for novelty runs dry or their value diminishes in other ways, e.g.~by becoming  obsolete or unreliable. They can then dis-intensify old ties and establish new ones, accumulate specializations over time, and alternate or combine specializing with brokering \cite{carnabuci09}. Both cross-sectionally and longitudinally, it is beneficial for them to vary the strength of their ties.

 Seen from a historical perspective, it is clear that since the beginning of industrialization, the potential benefits of combining information have increased considerably \cite{mokyr02}, if irregularly and unequally. The importance of varying tie strength has increased accordingly.

\section*{Appendix: a mini course in brokerage}
To understand the network structure of brokerage in general, refer to Figure~1. For simplicity's sake, information is transmitted in both directions for all pairs of actors, so the ties are edges. Obviously, a static network image is an idealization of  a far more complex pattern of social interactions, where the transmission of information is neither continuous nor simultaneous. Yet this simplification works quite well in many cases. In Figure~1,  focal node $A$ may combine different information received from mutually disconnected $C$, $D$ and $E$, which would be unproblematic for $A$ if $B$ were not there. If $B$ uses and recombines information received from $C$ and $D$, for example $C$'s demand and $D$'s offer in a market, $B$ competes with $A$ and reduces $A$'s chance to strike a deal first. In creative fields, $B$'s resulting idea is likely to be more similar to $A$'s, who draws partly from the same sources, than if $B$ were to use sources unrelated to $A$'s; $B$ does not necessarily reduce $A$'s information diversity, but it does reduce $A$'s chance for novelty \cite{carnabuci10}. In general, \emph{structurally equivalent} others, who are to some extent related to the same alters as ego (i.e.~have niche overlap), tend to reduce ego's opportunities, and reduce the value of the structural hole for ego. 
Let us now examine the opposite case, when $B$ provides useful information to $C$ and $D$; they get the benefit first, and may increase their benefit by cross-fertilizing $B$'s idea with $A$'s. In this situation, $A$ has a timing disadvantage, might receive $B$'s idea incomplete or distorted through $C$ or $D$, or may not hear about $B$'s idea at all. The latter is almost certainly the case for ideas that $B$ receives from $F$. The example illustrates that $A$ should be in direct contact with her sources; in this case timing and reliability are best. Information is useful to a broker where and when the news breaks \cite{burt92}, while ``second-hand brokerage" is not \cite{burt07}, hence long paths in the widely used measure of betweenness are to be truncated to paths of length two, resulting in what might be called between-two-ness (Eq.~4). 
Finally, alters connected directly, here by adding a tie between $C$ and $D$, may exchange information without involving $A$ at all, thereby removing $A$'s brokerage opportunity entirely. In sum, useful nodes for ego are at a distance of two ties away from each other, such that ego sits astride between those places where ``useful bits of information are likely to air, and provide a reliable flow of information to and from those places" \cite{burt92}. 

For focal node $i$ and its contact nodes $j$ and $l$, a structural hole is the lack of a direct tie between $j$ and $l$; then the numerator (Eq.~4) of between-two-ness $g_{jil} = 1$. If there is a tie between $j$ and $l$, there is no structural hole, and $g_{jil} = 0$. If there is a structural hole, its value diminishes with the number of nodes that access it, as discussed above, which is expressed in the denominator in Eq.~4, where $i$'s competitors are indexed on the dot in $g_{j.l}$. In Figure~1, for the structural hole between $C$ and $D$, $g_{C.D} = 2$, because both $A$ and $B$ access it. 
 To compute between-two-ness of focal node $i$, one simply adds up its competition-weighted structural holes across all pairs of contact nodes, 

\begin{equation}
     C_{B2}(i) = \sum_{j<l} \frac{g_{jil}}{g_{j.l}}
    \label{betweenness}
    \end{equation} 
$A$ brokers three structural holes of which one is contested by $B$, and accordingly, $C_{B2}(A) = 2.5$. Notice that if ties are asymmetric, as they are in citation networks, a path in one direction of the arcs is often different, and is to be counted separately, from a path in the opposite direction.
\begin{figure}
 \begin{center}
\usetikzlibrary{positioning}
\begin{tikzpicture}[xscale=6,
     yscale=8,>=stealth]
  \tikzstyle{v}=[circle,
     minimum size=1mm,draw,thick]
  \node[v] (a) {$\mathbf{A}$};
  \node[v] (b) [right=of a] {$C$};
  \node[v] (c) [below=of a] {$D$};
  \node[v] (d) [below=of b] {$B$};
  \node[v] (e) [left=of a] {$E$};
  \node[v] (f) [right=of d] {$F$};
  \draw[thick,-]
        (a) to node {} (c);
  \draw[thick,-]
        (a) to node {} (e);
  \draw[thick,-]
        (d) to node {} (f);
  \draw[thick,-]
        (a) to node {} (b);
  \draw[thick,-]
        (c) to node {} (d);
  \draw[thick,-]
        (b) to node {} (d);
\end{tikzpicture}
  \caption{\small A network wherein focal node $A$ brokers three structural holes between her contacts $E$, $C$, and $D$, while $B$ competes for the hole between $C$ and $D$.}
  \end{center}
 \end{figure}
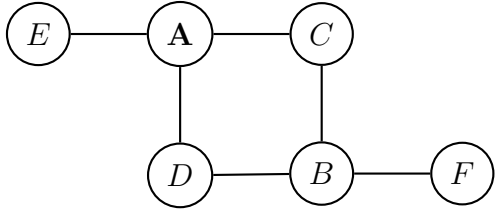

\end{document}